\documentclass[twocolumn,epjc3]{svjour3} 
\usepackage{amssymb,amsfonts,amsmath,mathtools,bm,color}

\newcommand{\be}{\begin{equation}}
\newcommand{\ee}{\end{equation}}
\newcommand{\bea}{\begin{eqnarray}}
\newcommand{\eea}{\end{eqnarray}}
\newcommand{\vep}{{\bm p}}
\newcommand{\vek}{{\bm k}}
\newcommand{\veq}{{\bm q}}
\newcommand{\vex}{{\bm x}}
\newcommand{\nn}{\nonumber}
\newcommand{\Hb}{\bar{H}}

\smartqed 
\RequirePackage{mathptmx} 
\RequirePackage[numbers,sort&compress]{natbib}
\journalname{Eur. Phys. J. C}

\begin{document}

\title{Remarks on the Heavy-Quark Flavour Symmetry for doubly heavy hadronic molecules}

\author{V. Baru\thanksref{e1,addr1,addr2,addr3}
 \and
 E. Epelbaum\thanksref{e2,addr4}
 \and
 J. Gegelia\thanksref{e3,addr5,addr6}
 \and
 C. Hanhart\thanksref{e4,addr5}
 \and
 U.-G. Mei{\ss}ner\thanksref{e5,addr1,addr5,addr6}
 \and
 A.V.~Nefediev\thanksref{e6,addr3,addr7,addr8}
}

\thankstext{e1}{e-mail: vadimb@tp2.rub.de}
\thankstext{e2}{e-mail: evgeny.epelbaum@ruhr-uni-bochum.de}
\thankstext{e3}{e-mail: j.gegelia@fz-juelich.de}
\thankstext{e4}{e-mail: c.hanhart@fz-juelich.de}
\thankstext{e5}{e-mail: meissner@hiskp.uni-bonn.de}
\thankstext{e6}{e-mail: nefediev@lebedev.ru}

\institute{Helmholtz-Institut f\"ur Strahlen- und Kernphysik and Bethe Center for Theoretical Physics, Universit\"at Bonn, D-53115 Bonn, Germany \label{addr1}
\and
Institute for Theoretical and Experimental Physics, B. Cheremushkinskaya 25, 117218 Moscow, Russia \label{addr2}
\and
P.N. Lebedev Physical Institute of the Russian Academy of Sciences, 119991, Leninskiy Prospect 53, Moscow, Russia \label{addr3} 
\and
Ruhr-Universit\"at Bochum, Fakult\"at f\"ur Physik und Astronomie, Institut f\"ur Theoretische Physik II, D-44780 Bochum, Germany\label{addr4} 
\and
Forschungszentrum J\"ulich, Institute for Advanced Simulation, Institut f\"ur Kernphysik and
J\"ulich Center for Hadron Physics, D-52425 J\"ulich, Germany \label{addr5} 
\and
Tbilisi State University, 0186 Tbilisi, Georgia \label{addr6} 
\and
Moscow Institute of Physics and Technology, 141700, Institutsky lane 9, Dolgoprudny, Moscow Region, Russia \label{addr7} 
\and
National Research Nuclear University MEPhI, 115409, Kashirskoe highway 31, Moscow, Russia \label{addr8}
}

\date{}

\maketitle

\begin{abstract}
The possibility for a common effective field theory for hadronic molecules with different heavy-qu\-ark
fla\-vo\-urs is examined critically. 
It is argued that such a theory does not allow one to draw definite conclusions for 
doubly heavy molecules. 
In particular, it does not allow one to relate binding energies for the molecules in
the $c$-quark and $b$-quark sectors with controlled uncertainties.
Therefore, while this kind of reasoning does not preclude from employing 
heavy-quark spin symmetry for charmonium- and bottomonium-like states separately within a well
established effective field theory framework, 
relations between different heavy-quark sectors can only 
be obtained using phenomenological approaches with uncontrolled uncertainties. 
\keywords{strong interactions \and effective field theory \and heavy quark flavours \and exotic hadrons}
\end{abstract}

\section{Introduction}

In the last decade, lots of states were found experimentally in the heavy quarkonium mass range that did
not at all fit into the scheme predicted by the until then very successful constituent quark model, for
recent reviews see, for example, 
Refs.~\cite{Lebed:2016hpi,Esposito:2016noz,Ali:2017jda,Olsen:2017bmm}. 
By now there is already a sizable number of states discovered in the charmonium and bottomonium mass
range that seem to qualify 
as such exotic hadrons. In particular, a non-$Q\bar Q$ nature is most apparent for the charged
resonances decaying into final states that
contain a heavy quark $Q$ and a heavy antiquark $\bar Q$, so that they must contain at least four
quarks in total. 

In order to understand the nature of the mentioned exotic states, it appears necessary to perform
studies with a sound connection to QCD.
This can be done by either using lattice QCD or effective field theories derived from QCD. The latter
allow one to make 
predictions based on the symmetries of QCD both exact and approximate. In the heavy-quark
sector the most natural ones to exploit are the Heavy Quark Spin Symmetry (HQSS) and the Heavy Quark
Flavour Symmetry (HQFS).
The former symmetry acts within each heavy-quark sector individually and relates exotic states with
various quantum numbers which differ from each other by the 
coupling of the light-quark cloud with the spin of a given heavy quark. The corresponding states
are called the spin partners. The latter symmetry is expected to relate various properties of exotic 
states containing different heavy quarks. Clearly, manifestations of the aforementioned symmetries
depend crucially on 
the studied system at hand. For example, both symmetries are known to be operative in quark-antiquark
systems, where the degrees of freedom related to the heavy quark can be integrated out in a 
controlled way, see, for example, the textbook treatment in Ref.~\cite{Manohar:2000dt}. 

In this paper we demonstrate that, while HQSS allows one to construct sensible EFTs for hadronic mo\-le\-cu\-les
form\-ed by two heavy open-flavour mesons and thus to fully control the uncertainties, an 
analogous EFT based on HQFS does not exist.\footnote{As will become clear below the same
argument does not apply to bound systems of light and heavy mesons. For those states it is possible to
relate the bottom and the charm sector (see Ref.~\cite{Guo:2017jvc} and references therein).}
Thus, one is left 
to rely on phenomenological estimates in order to get some insight into the flavour partner states. 
We argue, therefore, that various calculations, which rely on HQFS (see, for 
example, Refs.~\cite{Nieves:2011vw,Guo:2013sya}), are phenomenologically motivated and have some merit,
however, they do not qualify as EFT calculations.

\section{Formalism and formulation of the problem}
\label{formalism}

The QCD Lagrangian is known to possess a well-defined heavy-quark limit \cite{Manohar:2000dt}. 
This implies that such a limit also exists for a system containing one heavy quark, and an EFT can be 
established to describe such a system with the uncertainties being fully under control. In particular,
the action of the theory can be expanded in the inverse powers of the heavy mass $M$,
\be
S=S_0+\frac{\kappa}{M}S_1+\left(\frac{\kappa}{M}\right)^2S_2+\ldots,
\label{Sexp} 
\ee
where $\kappa$ stands for an intrinsic scale of the theory related to the light degrees of freedom. Obviously,
for a heavy--light system, the leading term in Eq.~(\ref{Sexp}) describes the light particle 
motion in the field of the static source, and a series of corrections to this limit can be
established with the help of a systematic $1/M$ expansion of the Lagrangian. A paradigmatic example of such 
an approach to QCD is given by the Heavy Quark Effective Theory, see, for example,
Refs.~\cite{Manohar:2000dt,Grozin:2004yc}. 
In particular, this implies that in the large-$M$ limit, both the action and the interaction potential
between the light and heavy quarks are $M$-independent in the leading order. 

A natural next step would be to extend these considerations to systems containing two heavy mesons. For 
definiteness, let us stick to pseudoscalar ($P$) and vector ($V$) meson ($D^{(*)}$ and $B^{(*)}$ in
the $c$- and $b$-sectors, respectively) which can be 
combined within a single nonrelativistic heavy meson (antimeson) superfield
$H_a = P_a+ V^i \sigma^i$ ($\bar{H}_a =\bar{P}_a - \bar{V}^i_a \sigma^i$), where 
$a$ is a $SU(2)$ isospin index. The interaction of such mesons at low energies can be
described with the nonrelativistic Lagrangian \cite{Mehen:2011yh}
\bea
{\cal L}&=&{\rm Tr}\left[H^\dagger_a \left(i \partial_0 +\frac{\bm\nabla^2}{2 M}\right)_{ba} H_b\right] 
+\frac{\Delta}{4}{\rm Tr}[H^\dagger_a\sigma^iH_a\sigma^i]\nonumber\\
&+&{\rm Tr}\left[\Hb^\dagger_a \left(i\partial_0+\frac{\bm\nabla^2}{2M}\right)_{ab}\Hb_b\right]+\frac{\Delta}{4}
{\rm Tr}[\Hb^\dagger_a\sigma^i\Hb_a \sigma^i]\nn\\[-2mm]
\label{Lag}\\[-2mm]
&-&\frac{C_{00}}{4}{\rm Tr}[\bar{H}_a^\dagger H_a^\dagger H_b\bar{H}_b] 
-\frac{C_{01}}{4}{\rm Tr}[\bar{H}^\dagger_a \sigma^i H^\dagger_a H_b \sigma^i \bar{H}_b ]\nonumber\\
&-&\frac{C_{10}}{4}{\rm Tr}[\bar{H}^\dagger_a \tau^A_{aa^\prime} H^\dagger_{a^\prime} H_b \tau_{bb^\prime}^A \bar{H}_{b^\prime} ]\nonumber\\
&-&\frac{C_{11}}{4}{\rm Tr}[\bar{H}^\dagger_a \tau^A_{aa^\prime} \sigma^i H^\dagger_{a^\prime} H_b \tau_{bb^\prime}^A
\sigma^i \bar{H}_{b^\prime}],\nn
\eea
where $\Delta=M_V-M_P$ and, for convenience, the isospin matrices are normalised as $\tau^A_{ab} \tau^B_{ba} =
\delta^{AB}$.
The transformation properties of the superfields $H_a$ and $\bar{H}_a$ under heavy quark spin and other
symmetries are given in Ref.~\cite{Fleming:2008yn}.
The first four terms in Eq.~(\ref{Lag}) are the leading heavy and anti-heavy hadron chiral perturbation
theory Lagrangian of Refs.~\cite{Wise:1992hn,Grinstein:1992qt,Burdman:1992gh,Yan:1992gz,Casalbuoni:1992dx}
written in the two-component notation \cite{Hu:2005gf}. The heavy mesons and anti-heavy 
mesons interact via the four remaining terms in the Lagrangian which describe $S$-wave contact interactions.
Contact interactions of this type were first written down in 
Ref.~\cite{AlFiky:2005jd}. The mass $M$ in the kinetic terms is the spin-averaged heavy-meson mass, $M=(3M_V+M_P)/4$. 

The Lagrangian (\ref{Lag}) allows one to establish a well-defin\-ed EFT in a given heavy-quark sector,
that is for a given fixed heavy mass $M$. The problem addressed in this paper is whe\-ther or not
\emph{one and the same EFT can be used to relate different heavy-quark sectors, that is treating $M$ 
as a parameter}.

\section{Attempts to construct a heavy-flavour EFT}

\subsection{Mass-independent action}
\label{sec:massindep}

Following Refs.~\cite{Luke:1996hj,AlFiky:2005jd},
we start assuming that \emph{there exists a finite limit for the action as $M\to\infty$}, and
the corrections are given as an expansion in the inverse powers of $M$, see Eq.~(\ref{Sexp}) 
above. 

In general, a system of two interacting heavy mesons has to be described by a complete set of
various coupled partial
waves as soon as one-pion exchange (OPE) is considered~\cite{Baru:2016iwj}. However, for the sake of simplicity,
we present our argument based solely on a single-channel calculation with $S$ waves only
within a particular channel. As a consequence of this, from the various parameters $C_{ij}$
that appear in the Lagrangian of Eq.~(\ref{Lag}) only some fixed linear combination
will control the system of interest --- this parameter we will generically call $C_0$ below.

We recall now that the power counting in the systems containing two heavy particles is quite nontrivial
and substantially different from that in the 
processes involving a single heavy particle. The reason for that is the presence of
pinch singularities which show up in the loop contributions
when, in the free interaction term $i \partial_0+{\bm\nabla^2}/(2M)$ in the effective Lagrangian,
the contribution of the temporal derivative to the scattering amplitude is taken to be 
of the leading order, and the contribution of the spatial part is considered as a higher order term
\cite{AlFiky:2005jd}. 
This problem is completely analogous to the one discussed extensively in the context of chiral EFT
for the two-nucleon system~\cite{Weinberg:1991um}. To cure this problem, 
the term ${\bm\nabla}^2/(2M)$ has to be included in leading order calculations of the observables
(together with $\partial_0$), as shown by Weinberg in Ref.~\cite{Weinberg:1991um}.
Following this logic, one might argue that its contribution in the action is non-vanishing
in the $M\to\infty$ limit alongside of the contribution of the time-derivative term.
Attempting to justify such an argument we rescale the time, the spatial coordinates and the field as
\be
t\to\xi_t t',\quad {\bm x}\to\xi_x{\bm x}',\quad H\to \xi_H H'.
\ee
Then the transformation of the action reads 
\bea
&&\int d^3 x dt\; {\cal L}(t,x,H) \nonumber\\
&&\hspace*{0.05\textwidth}=\int d^3 x' dt'\; (\xi_x^3 \xi_t{\cal L}(\xi_t t',\xi_x x',\xi_H H')) \label{scale1}\\
&&\hspace*{0.2\textwidth}=\int d^3 x' dt' {\cal L}'(t',x',H'), \nonumber
\eea
and it is assumed that there exists a set of transformations $\{\xi_t,\xi_x,\xi_H\}$ such that
the mass $M$ drops out explicitly from the action (up to the corrections suppressed in the limit 
$M\to\infty$). 

Considering the free part of the Lagrangian (\ref{Lag}) and demanding that both the temporal
derivative $\partial_0$ and the term $\bm\nabla^2/(2M)$ appear at 
the same order in $M$ not only in the observables to avoid pinch singularities but also in the action,
it is easy to find that we need to take
\be
\xi_t=M\xi_x^2,\quad\xi_H=\xi_x^{-3/2}.
\ee
Then, indeed, the kinetic term in the rescaled Lag\-ran\-gi\-an ${\cal L'}$ takes the form
\be
i \partial_0' +\frac12{\bm\nabla}'^2,
\label{kin2}
\ee
where the heavy mass has disappeared and the two terms in Eq.~(\ref{kin2}) appear to be of the same order. 
As a consequence of this rescaling, 
the coefficient in front of the $H^{\prime\dagger}\bar{H}^{\prime\dagger}H'\bar{H}'$ interaction term becomes
$M C_0/\xi_x$ where, as already stated above, for simplicity, we do not distinguish between
different contact terms in 
Eq.~(\ref{Lag}) and use the notation $C_0$ for all of them.\footnote{We recall that, 
in the strict heavy-quark limit of $M\to\infty$, the HQSS violating terms $\propto\Delta$
in the Lagrangian Eq.~(\ref{Lag}) should be disregarded.} 

It is, therefore, sufficient to demand that 
\be
C_0=\frac{\xi_x C_0^{(1)}}{M}+{\cal O}\left(\frac{\xi_x}{M^2}\right),
\label{C0b}
\ee
with a mass-independent coefficient $C_0^{(1)}$ to ensure the existence of a finite limit of the action
as $M\to\infty$. 

It has to be noticed, however, that the interaction between the heavy mesons is not exhausted by the
short-range potential $C_0$ but there is also a contribution from the light-meson exchanges, the pion exchange 
being the most prominent example of such an interaction. The free and interacting parts of the pionic 
Lagrangian read
\bea
{\cal L}_{\pi}&=&\frac12 \pi^\dagger\left ( {\partial_0}^2 - {{\bm\nabla}^2}- m_{\pi}^2 \right ) \pi\nn\\[-2mm]
\label{Lpi}\\[-2mm]
&+& \frac{g}{f_{\pi}} \left({\rm Tr}[ {H}_a^\dagger H_b \sigma^i]-{\rm Tr}[\bar{H}_a \bar{H}_b^\dagger
 \sigma^i]\right) \nabla \pi^i_{ab},\nn
\eea
where $g$ denotes the coupling constant of the heavy--light mesons with the pion and $f_{\pi}$ is the
pion decay constant. 
One has to require, therefore, that taking the limit $M\to\infty$ in the action leads to a finite
result \emph{including the pionic part} specified in Eq.~(\ref{Lpi}). 
Then, after rescaling, the free part of the pionic Lagrangian becomes
\be
{\cal L}_\pi'=\frac12\xi_\pi^2\pi'^\dagger\left(\frac{1}{M^2\xi_x^4}{\partial'_0}^2-
\frac{1}{\xi_x^2}{\bm\nabla}'^2-m_{\pi}^2\right) \pi', 
\label{Lpifree}
\ee
where we rescaled the pion field as $\pi\to \xi_{\pi} \pi'$. For the action this yields
\begin{equation}
\frac12\int dt'd^3 \vec{x}'(M\xi_x\xi_\pi^2)\pi'^\dagger\left(\frac{1}{M^2}{\partial'_0}^2
-\xi_x^2{\bm\nabla}'^2-\xi_x^4 m_{\pi}^2\right)\pi'.
\label{action0}
\end{equation}
The action Eq.~(\ref{action0}) has a physically adequate finite limit as $M\to\infty$ 
for\footnote{Remarkably, for $\xi_x=1$, the resulting leading mass dependence (see Eq.~(\ref{C0b})), 
$C_0\propto 1/M$, is identical to the one also found in Ref.~\cite{AlFiky:2005jd}.}
\be
\xi_x=1,\quad \xi_\pi=1/\sqrt{M}.
\ee
Indeed, in this case, in full agreement with natural expectations, the temporal derivative term
vanishes and the pion propagator in the momentum space takes its static form, 
\be
G_{\pi}({\bm q})= \frac{1}{{\bm q}^2 +m_\pi^2}.
\ee
This results in the standard static Yukawa-type interaction potential
between two static sources.

It is now straightforward to find the contribution to the action from the interaction part
of the Lagrangian of Eq.~(\ref{Lpi}),
\be
\int dt'd^3 \vex'\frac{g\sqrt{M}}{f_\pi}\left({\rm Tr}[ {H'}_a^\dagger H'_b \sigma^i]
-{\rm Tr}[\bar{H'}_a \bar{H'}_b^\dagger \sigma^i]\right) \nabla \pi'^i_{ab}, 
\label{Lpiint}
\ee
which blows up in the limit $M\to\infty$ because it is well-known that the pion coupling
constant to a heavy field is independent of the mass of this field up to the corrections proportional to inverse powers of the mass $M$. 
In particular, the pion coupling should survive in the limit $M\to\infty$ \cite{Manohar:2000dt,Wise:1992hn}.
Indeed, the value for the $D^*D\pi$ coupling constant extracted from the experimentally measured partial decay width $D^{*+}\to D^0\pi^{+}$ 
is consistent with that for the $B^*B\pi$ coupling constant from the recent lattice QCD analysis \cite{Bernardoni:2014kla}. 

Based on the consideration above, one is led to conclude that the hypothesis employed that the derivative
term ${\bm\nabla}^2/(2M)$ contributes to the leading $M$-in\-de\-pen\-dent part of the action, 
as was advocated in Ref.~\cite{AlFiky:2005jd}, cannot be
correct. Therefore, this term is of a subleading order in the $1/M$ expansion of the action, however,
it gives a leading-order contribution to the scattering amplitudes of the systems with two heavy 
mesons. Because the guiding principle discussed above appears to be contradictory,
the constraints on the mass dependence of the interaction $C_0$ drawn from it
can not be treated as reliable. 
While we still assume the action to possess the expected expansion (\ref{Sexp}), contributions to
various terms are generated by the effective Lagrangian
which contains terms with explicit factors of the inverse powers of the large mass $M$ as well as
an implicit dependence on this parameter through various low-energy constants. 

\subsection{Mass-independent contact interaction}
\label{massindep}

It was argued in Refs.~\cite{Nieves:2011vw,Guo:2013sya} that the heavy-mass limit in a doubly
heavy system implies that \emph{the low-energy constants are independent of the heavy mass}. As was 
explained above, one may consider 
this claim to be a natural extension of the results for a heavy--light system which follow directly
from the heavy-quark limit of the QCD Lagrangian \cite{Manohar:2000dt}. 
However, as is demonstrated below, in a low-energy theory with contact interactions only this
assumption does not lead to a well-defined EFT.

The Lippmann-Schwinger equation for the elastic scattering amplitude reads 
\be
T(E)=-C_0-C_0\Sigma(E)T(E),
\ee
with
\be
\Sigma(E) = \int \frac{d^3 q}{(2\pi)^3}G(q,E),
\ee
\be
G(q,E)=\frac{1}{E-q^2/(2\mu)+i0},
\label{GqE}
\ee
where $\mu=M_1M_2/(M_1+M_2)=M/2$ is the reduced mass of the two mesons
where the latter identity holds for $M_1=M_2=M$ that we will assume for simplicity below. 
Note that it is the kinetic energy of the two-meson system that provides the 
right-hand cut which introduces the most relevant heavy meson mass dependence.
In particular, for a heavy--light system one finds that $\lim_{M_2\to \infty}\mu = M_1$
stays finite and accordingly such systems have a controlled heavy quark
mass limit, which in turn allows, for example, for controlled predictions for
hadronic molecules in the $b$ sector from a theory with parameters fixed in
the $c$ sector --- this kind of studies was pioneered in Ref.~\cite{Kolomeitsev:2003ac}; for
a discussion of the heavy quark limit of the loop function of heavy--light systems see, for example,
Refs.~\cite{Cleven:2010aw,Altenbuchinger:2013vwa} or the recent review of Ref.~\cite{Guo:2017jvc}.
However,
 if both masses go to infinity the reduced mass goes to infinity as 
well. It is this fact that drives the difference between heavy--light and
heavy--heavy systems mentioned above.

The loop integral $\Sigma(E)$ is linearly divergent and needs to be 
regularised. Since 
in a well-defined EFT the final result cannot depend
on the regularisation procedure, we stick to the simplest, sharp cut-off scheme to arrive at
\begin{equation}
\Sigma(E)=-\frac{M}{4\pi}(\tilde{\Lambda}+ik),\quad \tilde{\Lambda}=\frac{2}{\pi}\Lambda,
\end{equation}
where $\Lambda$ is a momentum UV regulator and $k=\sqrt{M E}$ is the on-shell momentum. The
same result can be obtained straightforwardly, for example, in the PDS sche\-me \cite{Kaplan:1998tg}, 
with $\tilde{\Lambda}$ substituted by the subtraction point.

If there exists a bound state, then, at the bound state pole, $T^{-1}(-E_B)=0$, so that 
\begin{equation}
0=-C_0^{-1}-\Sigma(-E_B)=-C_0^{-1}+\frac{M}{4\pi}(\tilde{\Lambda}-\gamma),
\label{RGeq}
\end{equation}
where we defined the binding momentum $\gamma=\sqrt{M E_B}$.

Under the assumption of the $M$-independence of the contact interaction $C_0$, Eq.~(\ref{RGeq})
relates the binding momentum in the $b$-sector, $\gamma_b$, with that in the $c$-sector, $\gamma_c$, as
\begin{equation}
M_c(\tilde{\Lambda}_c-\gamma_c)=M_b(\tilde{\Lambda}_b-\gamma_b),
\label{gbgc}
\end{equation}
that is
\begin{equation}
\gamma_b=\frac{M_c}{M_b}\gamma_c-\frac{M_c}{M_b}\tilde{\Lambda}_c+\tilde{\Lambda}_b=: \frac{M_c}{M_b}\gamma_c+\Lambda_{\rm UV},
\label{gammadep}
\end{equation}
where $\Lambda_{\rm UV}=:\tilde{\Lambda}_b-(M_c/M_b)\tilde{\Lambda}_c$ and
$M_c$ ($M_b$) denotes the hadronic mass in the $c$-quark ($b$-quark) sector. Assuming that\footnote{Phenomenologically this
assumption is well motivated, since the cut-off basically controls the light quark dynamics
which is expected to be independent of the heavy quark flavour.} $\tilde{\Lambda}_c\simeq\tilde{\Lambda}_b=\tilde{\Lambda}$
and using the $X(3872)$ state treated as a $D\bar{D}^*$ bound state as input,
the authors of Refs.~\cite{Nieves:2011vw,Guo:2013sya} employed 
the relation~(\ref{gammadep}) to predict the existence of a $1^{++}$ bound state
$X_b$ near the $B\bar{B}^*$ threshold. 
It is easy to see then that 
\be
\gamma_b=\gamma_c+\left(1-\frac{M_c}{M_b}\right)(\tilde{\Lambda}-\gamma_c)\simeq \frac4{3\pi}{\Lambda},
\label{c2b}
\ee
where it was used that $M_b\approx 3M_c$ and $\gamma_c\ll\Lambda$. For $\Lambda\simeq 500$~MeV,
the binding energy of the hypothetical $X_b$ 
state, $E_B=\gamma_b^2/M_b$, takes values of the order of 10~MeV in agreement with the
findings of Refs.~\cite{Nieves:2011vw,Guo:2013sya}. In other words, a heavier $\bar{b}b$ system appears to 
be stronger bound than a lighter $\bar{c}c$ system. 

It has to be noticed, however, that, for a given fixed value of the binding momentum in the $c$-sector,
the binding momentum in the $b$-sector is entirely controlled by the 
UV regulator $\Lambda$, see Eq.~(\ref{c2b}). From the EFT point of view, this means that a HFS
violating contact interaction is required already at leading order to absorb the dependence of the 
observables on the regulator. This, however, contradicts the assumption that the leading-order
contact interaction $C_0$ is mass-independent.

We conclude, therefore, that the assumption of a mass-independence of the contact interaction
does not lead to an EFT that is renormalisable.

\subsection{Renormalisibility}

It was demonstrated in the previous subsections that none of the two assumptions, 
which seem to follow naturally from the success of building an EFT in a
heavy--light system allows one to develop a self-consistent EFT in a heavy--heavy system. 
In this subsection we impose the most general condition of renormalisability on such an EFT under 
construction and consider consequences of such a setup.

Equation (\ref{RGeq}) is the starting point of our investigation. 
As an observable quantity $\gamma$ cannot depend on the UV regulator, the entire $\Lambda$-dependence
has to be driven by the unobservable short-range potential $C_0$, that gives
\begin{equation}
C_0^{-1}(\Lambda,M)=\frac{M}{4\pi}(\tilde{\Lambda}-\gamma)=:\frac{M}{4\pi}\tilde{\Lambda}+C_{0R}^{-1}(M),
\label{CR}
\end{equation}
in line with the findings of Ref.~\cite{Kaplan:1998tg}, where, alternatively, the coupling was fixed to
the scattering length instead of the binding momentum. The quantity $C_{0R}$, defined
through Eq.~(\ref{CR}), is the renormalised,
regulator-independent contact term which describes short-range interactions in the system. It is 
trivially related with the binding energy $\gamma(M)$,
\be
\gamma(M)=-\frac{4\pi}{M C_{0R}(M)}.
\label{gammaCR}
\ee

It is instructive to solve relation~(\ref{CR}) for $C_0(\Lambda,M)$,
\begin{equation}
C_0(\Lambda,M)=\frac{C_{0R}(M)}{1+M\tilde{\Lambda}C_{0R}(M)/(4\pi)},
\label{C0e}
\end{equation}
and investigate a possible dependence of the bare contact term $C_0$ as a function of the mass $M$
in the limit $M\to\infty$. Suppose that 
\be
C_{0R}(M)\propto 1/M^n
\label{C0Ralpha}
\ee
in this limit, with $n$ taking any value from $-\infty$ to $\infty$. Then
\be
C_0(\Lambda,M)\mathop{\propto}_{M\to\infty}
\left\{
\begin{tabular}{lcc}
$1/M^n$,&&$n>1$,\\
$1/M$,&&$-\infty<n\leqslant 1$,
\end{tabular}
\right.
\label{C0behaviour}
\ee
that is, renormalisability of the contact theory requires that $C_0$ decreases at least as $1/M$
in the limit $M\to\infty$. 

From these considerations we conclude that in general the contact interaction has the form 
\be
C_0(\Lambda,M)=\sum_{n=0}^\infty \frac{C_0^{(n)}(\Lambda)}{M^n},
\label{C0General}
\ee
with $C_0^{(0)}=0$. Furthermore, none of our general arguments above suggests that a few more
coefficients $C_0^{(n)}$ cannot vanish as well. 
Then, as follows from Eq.~(\ref{C0behaviour}), any behaviour of $C_{0R}(M)$ for large values
of $M$ is compatible with Eq.~(\ref{C0General}).
Therefore, within the renormalisable pionless EFT no constraint is imposed on
the behaviour of the renormalised contact term $C_{0R}(M)$. Equivalently, renormalisability of the
theory alone does not fix the $M$-behaviour of the 
binding momentum (\ref{gammaCR}), and the corresponding EFT lacks predictive power.

\subsection{EFT with pions}

In this chapter we proceed to include one-pion-exchange (OPE) on top of the contact interaction to check
whe\-ther the conclusions arrived in the previous subsection persist.

As seen from Eq.~(\ref{C0e}), in the pionless renormalisable EFT from the previous chapter,
the bare coupling $C_0$ is suppressed in the limit $M\to\infty$, see Eq.~(\ref{C0behaviour}). 
Inclusion of pions 
changes the simple behaviour (\ref{C0e}) turning it to a more complicated one. However, for a
vanishing coupling of the pions to heavy mesons $g$, the formulae for the purely contact interactions 
alone have to be restored. 
Therefore, if the $M$-dependence of the renormalised contact interaction $C_{0R}$ cannot be
fixed from general principles in the pionless theory, it cannot be fixed in a theory with pions 
either.
Let us illustrate this argument by considering a simple case of the uncoupled $S$-wave
scattering by taking LO potential as a contact interaction plus OPE.
We start by reminding the reader that the OPE potential provides not only a long-range
contribution to the heavy meson-antimeson interaction at large distances but also contains a short-range 
part, that is well defined in the sense of an EFT only in connection with a contact operator
\cite{Baru:2015nea}. In particular, for the $V\to P\pi$ vertex (for 
example, for the $D^*\to D\pi$ or $B^*\to B\pi$ one) the static OPE potential behaves as
\be
V_{\rm{OPE}}\propto \frac{({\bm\varepsilon}_1{\bm q})({\bm\varepsilon}_2^*{\bm q})}{{\bm q}^2+m_\pi^2},
\label{VOPE}
\ee
where the ${\bm\varepsilon}$ are polarisation vectors of the initial- and final-state vector mesons
and ${\bm q}$ is the pion momentum. 
After partial wave decomposition, the $S$-wave part of this 
 potential tends to a constant in 
the limit $q\to\infty$ which is an indication of a short-range dynamics contained in the OPE. 

As mentioned above, we consider the LO potential as a contact interaction plus OPE,
\begin{eqnarray}
V(\vep,\vep')=C_0+\frac{\alpha \veq^2}{\veq^2+m_\pi^2}&=&(C_0+\alpha)-\frac{\alpha m_\pi^2}{\veq^2+m_\pi^2}
\nonumber\\[-1mm]
\label{LOPOT}\\[-1mm]
&=:&C_0'+V_\pi (\vep,\vep'),\nonumber
\end{eqnarray}
with $\veq=\vep-\vep'$ and $\alpha$ a constant which depends on the particular 
partial wave and spins of the heavy meson pair. We solve the Lippmann--Schwinger equation, 
\begin{equation}
T(\vep,\vep')=V(\vep,\vep')+M\int\frac{d^3k}{(2\pi)^3}\frac{V(\vep,\vek)T(\vek,\vep')}{k^2-q^2+i\epsilon},
\label{LOT}
\end{equation}
to arrive at \cite{Kaplan:1996xu}
\begin{equation}
T(\vep,\vep')=T_\pi(\vep,\vep')+X(\vep)\tau(M)X(\vep'),
\label{LOTSol}
\end{equation}
where 
\bea\label{solution}\\[-1mm]
\tau^{-1}(M)=C_0^{\prime-1}
&-&M\int\frac{d^3k}{(2\pi)^3}\frac{X(\vek)}{k^2-q^2+i\epsilon}~.\nonumber
\eea
$T_\pi(\vep,\veq)$ is the solution of Eq.~(\ref{LOT}) for $V(\vep,\vep')=V_\pi(\vep,\vep')$ and 
\begin{equation}
X(\vep)=1+M\int\frac{d^3k}{(2\pi)^3}\frac{T_\pi(\vek,\vep)}{k^2-q^2+i\epsilon}~.
\label{defchi} 
\end{equation}

Part of the integral in Eq.~(\ref{solution}) corresponding to the second term in Eq.~(\ref{defchi})
contains a logarithmically divergent contribution which is to be absorbed by a higher order 
$m_\pi$-dependent 
counter term. In the meantime, the integral corresponding to the first term in Eq.~(\ref{defchi})
is linearly divergent, and this divergence is removed by renormalising 
$C_0'$ as (\emph{cf.} Eq.~(\ref{CR}))
\begin{equation}
C_0^{\prime-1}(\Lambda,M)=\frac{M}{4\pi}\tilde{\Lambda}+C_{0R}^{\prime-1}(M)~.
\label{CtildeR}
\end{equation}
Therefore,
\begin{equation}
C_0(\Lambda,M)=-\alpha+\frac{C_{0R}'(M)}{1+M\tilde{\Lambda}C_{0R}'(M)/(4\pi)},
\label{C0eT}
\end{equation}
which is fully analogous to the relation~(\ref{C0e}). Therefore, the analysis performed
after Eq.~(\ref{C0e}) applies to Eq.~(\ref{C0eT}) as well and so does the conclusion that
the structure of $C_0$ specified in Eq.~(\ref{C0General}) does not restrict the $M$-dependence 
of $C_{0R}'(M)$, thus leaving the $M$-dependence of the pole position of the amplitude completely
uncontrolled. 

In other words, the inclusion of pionic degrees of freedom does not change the conclusion of
the previous subsection that it appears to be not possible to construct a heavy-flavour EFT
that could relate observables in the $c$-quark and $b$-quark sectors with controlled uncertainties.

\section{Summary and discussion}

To summarise the results presented in this work, one is led to conclude that
a properly renormalisable pionless EFT that relates different heavy quark sectors
needs to be built from bare contact interactions which scale as powers of $1/M$, whe\-re
$M$ is proportional to the heavy-meson mass. The same mass
scaling was already proposed in Ref.~\cite{AlFiky:2005jd} based on demanding
a proper power counting for a heavy--heavy system. 
However, the heavy-mass-de\-pen\-den\-ce of the re\-nor\-ma\-lis\-ed contact terms remains completely
unfixed and, therefore, so does the heavy-mass-dependence 
of the observable quantities. The inclusion of pions changes the heavy-mass-de\-pen\-den\-ce of the
bare strength of the contact
terms, however, still does not help to fix the mass-de\-pen\-den\-ce of the renormalised 
short-range potentials.
Thus, one is forced to conclude that no common EFT for heavy--heavy
molecular states can be built which respects heavy-quark flavour symmetry and allows one to
relate observables in the $c$-quark and $b$-quark sectors in a controllable way.
A renormalisable EFT can only be built within a given heavy-quark sector with a fixed
heavy mass exploiting, for example, heavy-quark spin symmetry.

Although it is known for a long time that heavy--heavy systems appear to be troublesome from the point of view of a proper definition of the large-$M$ limit for them (see, for example, 
warnings contained in Refs.~\cite{Jenkins:1992nb,Flynn:2007qt}), the present paper puts
this statement on a mathematically rigorous ground.
In particular, we show that for a heavy--heavy system general 
principles alone, like renormalisability and a proper scaling behaviour of the action with the heavy mass, 
do not allow one to build a common EFT relating the $c$- and $b$-quark sectors which 
allows for controlled uncertainty estimates. In other words, the 
existence of such a EFT would mean the possibility to relate observables in the two sectors with \emph{any} prescribed accuracy provided the theory is systematically considered to the necessary 
order, as this takes place, for example, for HQSS within a particular heavy-quark sector. It is important to emphasise that our findings do not have the form of a no-go theorem for relating 
observables in the 
charm and bottom sectors in general, however, to do so one needs to invoke additional assumptions and take care of the accuracy of the approximations made, since the latter are not controlled by 
a consistent power counting and as such may appear unreliable. 

Finally,
had we known the exact scattering potential including its short-distance (large-momentum) behaviour
as well as its heavy mass dependence, then we would have been able to determine the poles of
the amplitude in the complex energy plane exactly without further ado. 
The mentioned meson-meson potential would be characterised
by an intrinsic range, the latter then playing a role of the regulator $\Lambda_{\rm UV}$ that
appeared in Eq.~(\ref{gammadep}) 
and this scale would be treated as physical. 
In this case, it would be indeed possible to relate $\gamma_c$ and $\gamma_b$ within the same formalism.

\begin{acknowledgements}
The authors are grateful to F.-K. Guo, J.~Nieves, A.~Nogga, and M.~P.~Valderrama for useful discussions. 
This work was supported in part by the DFG (Grant No. TRR110) and the NSFC (Grant No. 11621131001) through the
funds provided to the Sino-German CRC 110 ``Symmetries and the Emergence of Structure
in QCD''. Work of V.B. and A.N. was supported by the Russian Science Foundation (Grant No. 18-12-00226).
J.G. acknowledges support from the Georgian Shota Rustaveli National Science Foundation (Grant No. FR17-354). 
The work of UGM was also supported by VolkswagenStiftung (Grant No. 93652) and by the Chinese Academy of
Sciences (CAS) President's International Fellowship Initiative (PIFI) (Grant No. 2018DM0034).
\end{acknowledgements}

\end{document}